\RequirePackage{fixltx2e}
\documentclass[prb,amsmath,superscriptaddress,twocolumn,showpacs]{revtex4}
\usepackage{bm}
\usepackage{amssymb}
\usepackage{colordvi}
\usepackage{color}
\usepackage{graphicx}
\usepackage{epstopdf}

\newcommand{\be}{\begin{equation}}
\newcommand{\ee}{\end{equation}}
\newcommand{\bea}{\begin{eqnarray}}
\newcommand{\eea}{\end{eqnarray}}
\newcommand{\bsube}{\begin{subequations}}
\newcommand{\esube}{\end{subequations}}

\begin{document}

\title{Temporal evolution of resonant transmission under telegraph noise}

\author{Shmuel Gurvitz}
\affiliation{Department of Particle Physics,
Weizmann Institute of Science, Rehovot 76100, Israel}
\affiliation{ Beijing Computational Science Research Center,
Beijing 100084, China}
\author{Amnon Aharony}
\affiliation{Department of Physics, Ben-Gurion University, Beer Sheva 84105, Israel}
\affiliation{Ilse Katz Center for Meso- and Nano-Scale Science and Technology,
Ben-Gurion University, Beer Sheva 84105, Israel}
\affiliation{Raymond and Beverly Sackler School of Physics and Astronomy, Tel Aviv University, Tel Aviv 69978, Israel}
\author{Ora Entin-Wohlman}
\affiliation{Department of Physics, Ben-Gurion University, Beer Sheva 84105, Israel}
\affiliation{Ilse Katz Center for Meso- and Nano-Scale Science and Technology,
Ben-Gurion University, Beer Sheva 84105, Israel}
\affiliation{Raymond and Beverly Sackler School of Physics and Astronomy, Tel Aviv University, Tel Aviv 69978, Israel}
\date{\today}

\begin{abstract}
The environment of a quantum dot, which is connected to two leads, is modeled by telegraph noise, i.e. random Markovian jumps of the (spinless) electron energy on the dot between two levels. 
The temporal evolutions of the charge on the dot and of the currents in the leads are studied using a recently developed single-particle basis approach, which is particularly convenient for the averaging over the histories of the noise. In the steady state limit we recover the Landauer formula. At a very fast jump rate between the two levels, the noise does not affect the transport. As the jump rate decreases, the effective average transmission crosses over from the transmission through a single (average) level to an incoherent sum of the transmissions through the two levels.
The transient temporal evolution towards the steady state is dominated by the displacement current at short times, and by the Landauer current at long times. It contains oscillating terms, which decay to zero faster than for the case without noise. When the average chemical potential on the leads equals the dot's ``original" energy, without the noise, the oscillations disappear completely and the transient evolution becomes independent of the noise.
\end{abstract}

\maketitle
\section{Introduction}

Quantum transport through mesoscopic systems\cite{Imry} 
is one of the most extensively investigated areas of theoretical physics. This transport is always affected by coupling to the environment. The  exact treatment of the whole system-environment complex represents a formidable many-body quantum mechanical problem. It is therefore useful to study approximate simple models, which may be justified in some special limits (see below) and which may capture the main aspects of the expected phenomena. One such approximation occurs if the back-action of the system on the environment can be neglected. In this case the action of the environment on the system can be treated as an external fluctuating time-dependent force. The entire many-body problem is then reduced to solving the Schr\"odinger equation with time-dependent potentials, which represent the fluctuating environment. Since the fluctuations of the environment are random, the final results typically ends up with some averaging over them.

After a long time, the system usually reached a steady state. 
In the presence of the above time-dependent random potentials, the steady state properties still require solving the time-dependent problem. In addition, the latter solution also yields the transient evolution of the system, from a specific initial configuration towards this steady state. These transient phenomena contain much information on the system and on the nature of its coupling to the environment. Experimentally, preparation of desired charge states in
quantum dots can be realized, and the availability of time-resolved measurement techniques
in nanoelectronics makes it feasible to reveal the special
roles played by the transient effects\cite{fuji}.

The time-dependent transport problem is not simple. Considerable efforts have been devoted to the development of time-dependent formulations for the transport, that capture its transient behavior. These include non-equilibrium Green �s function methods (NEGF) \cite{1}, master equations techniques\cite{114,2,weimin}, time-dependent density functional theory (TDDFT) \cite{3}, schemes that combine TDDFT and NEGF with the hierarchical equation of motion approach to open quantum dynamics \cite{4}, hybrid Floquet-NEGF treatments \cite{5} and time-dependent scattering state methods\cite{avi,tu}.
All of these schemes are quite heavy, and it is not easy to adapt them to the case of randomly fluctuating potentials. In the present paper we use the recently proposed single-electron approach (SEA)\cite{single}, based on the single-electron {\it Ansatz} for the many-electron wave-function. 
This approach solves the time-dependent Schr\"odinger equation for a single electron, yields simple expressions for the transient currents for time-dependent potentials, and reproduces the Landauer formula\cite{lan} for the steady state limit.  The SEA is a simple tool, which can be used to calculate the transient electric currents and charges in any mesoscopic system of coupled dots. Reference \onlinecite{single} demonstrated an application of the SEA to time-dependent transport in oscillating potentials and to electron pumping.

Since the SEA is valid for arbitrary time-dependent potentials, it can be used for randomly fluctuating potentials as well.
In this paper we combine the SEA approach with an appropriate averaging to study the effects of a simple fluctuating environment on the resonant transport through a single level quantum dot. As we show, this approach is particularly convenient for such averaging.

The simplest model for the environment is a single degree of freedom, called an ``elementary fluctuator", which fluctuates between two states.  The potential seen by the system then fluctuates randomly between two values. Such a random fluctuating potential is called ``telegraph noise" \cite{sd}.
Reference \onlinecite{amnon} listed several examples, in which back action from the system onto such a fluctuator could indeed be ignored.
In one example, a single impurity near the system is either empty or occupied
by a single electron (double occupancy is prevented by a strong Coulomb blockade)\cite{IT}. Another example concerned a two-level system at equilibrium with a heat bath\cite{13}. For both the single impurity and the two level system it has been argued\cite{alt} that back action can be ignored when the dynamics
of the fluctuating background charge or the two-level
system is governed by its coupling to a thermalizing heat
bath, which is much stronger than its coupling to the system.
The telegraph noise model is also justified in the limit of a
very high temperature of this heat bath\cite{18}.
In a third (non-equilibrium) example, current flows between two reservoirs through a single electron transistor (SET)
located near the system.  The states of the environment (SET plus reservoirs)
include states in which an arbitrary number of electrons
have moved between the two reservoirs, while the SET
can be (singly) occupied or empty.  When the bias
voltage between the reservoirs is much larger than any other energy in the
problem, Ref. \onlinecite{14} found that the Coulomb potential of the SET, as seen by the system, jumps randomly between its values when the SET is occupied or empty.

Here we assume that one can ignore back action, and describe the effects of the environment by the telegraph noise model. Even if this assumption is only approximate, we expect the results to
be qualitatively correct, and to stimulate future work which will test them. The advantage of our approach is obvious: it yields simple analytic expressions, which give insights into the way various physical quantities depend on the parameters.

\begin{figure}[tbh]
\includegraphics[width=7cm]{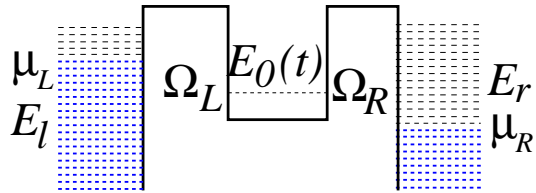}
\caption{(Color online) Resonant tunnelling through a single-level dot (see text)}.
\label{fig1}
\end{figure}

Most of the papers which use telegraph noise in mesoscopic systems have been motivated by quantum computing, and aimed to reduce the decoherence caused by the environment in closed systems, e.g. an {\em isolated} qubit \cite{amnon,IT,13,18,alt,19,14,alt2}.  Since environmental noise is very important in the dynamics of open quantum systems which are coupled to electronic reservoirs, the present paper presents the effects of telegraph noise on the simplest such system, i.e. resonant tunnelling through a single level of a quantum well (dot), coupled
to two reservoirs, as shown in Fig. \ref{fig1}. Some of our results in the steady state limit were already obtained twenty two years ago\cite{YG}, using a completely different approach, but we are not aware of other treatments of open systems with telegraph noise.   Here we use the SEA approach to obtain many new results, especially on the transient time evolution of the system towards the steady state. Unlike many of the treatments of closed systems, the SEA works with the wave functions, and not with the reduced density matrices. The approach can easily be extended to other open systems.

The system contains spinless electrons, so that the dot can only contain one electron or be empty. The environment is assumed to affect only the energy level of the electron on the dot,  which jumps randomly between the two values
$\overline E^{}_0\pm U/2$,
\begin{align}
E^{}_0(t)=\overline E^{}_0+\xi(t)U/2\ ,
\label{noise2}
\end{align}
where $\xi (t)=\pm 1$ is a random variable, which
follows a Markov process \cite{sd,amnon}: it jumps randomly from $1$ to $-1$ (and vice versa) at the same rate $\gamma /2$ (independently of its previous history).
 The two values of this energy represent the two states of the environment. For example, the energy $U$ can represent the Coulomb energy between the electron on the dot and the electron on the neighboring SET.

The SEA approach of Ref. \onlinecite{single} is summarized (with some extensions and some corrections) in Sec. II.
Section III uses this approach to calculate the currents and the charges on the leads and on the dot, ending up with the example of the time-independent case. The telegraph noise general theory, and its application to our model, are described in Sec. IV, while Secs. V and VI present the results for the steady state and for the transient currents and charges, respectively. Section VII presents our conclusions. The Appendix summarizes the Shapiro-Loginov formula, which is a very useful tool in deriving averages for the telegraph noise.

\section{Single-electron approach}

 The system in Fig. \ref{fig1} is described by
the following tunneling Hamiltonian:
\begin{align}
&H(t)=\sum_lE_l^{}a_l^\dagger a_l^{}+\sum_rE_r^{} a_r^\dagger a_r^{}+E_0(t)a_0^\dagger a_0^{}
\nonumber\\
&+\Big(\sum_{l}\Omega_L a^\dagger_l a_0^{} +\sum_{r}\Omega_R a^\dagger_r a_0^{}+H.c.\Big)\, ,
\label{a1}
\end{align}
where the operator $a_{l(r)}^{\dagger}$ creates an electron with energy $E^{}_{l(r)}$ in the left (right) reservoir and $a_{0}^{\dagger}$ is the same inside the quantum dot. The Hamiltonian $H$ is {\em time-dependent}, via an arbitrary time-dependence of the energy level $E_0^{}(t)$. We shall later discuss the special random case presented in Eq.~(\ref{noise2}).  The tunneling coupling energies of the reservoir states with the dot, $\Omega^{}_{L,R}$, are independent of $E^{}_{l,r}$ and of time and can always be taking real values(by an appropriate gauge transformation). Reference \onlinecite{single} analyzed the time-dependent couplings in the adiabatic limit.

At zero temperature, the initial state of the left and right reservoirs contains $N^{}_L$ and $N^{}_R$ electrons, respectively, which fill the single electron states with energies $\{E^{}_{\bar l}\le \mu^{}_L,\ E^{}_{\bar r}\le \mu^{}_R \}$, where $\mu^{}_{L(R)}$ is the Fermi energy in the left (right) reservoir. Concerning the quantum dot, we extend the discussion in Ref. \onlinecite{single} and consider two initial states, one in which the dot is occupied ($n^{}_0=1$) and one in which it is empty ($n^{}_0=0$).
The total number of electrons remains constant in time, $N+n^{}_0=N^{}_L+N^{}_R+n^{}_0$.
We thus start with the many-body pure state,
\begin{align}
|\Psi(0)\rangle =(a_0^\dagger)^{n^{}_0}\prod_{k}a_{k}^\dagger |0\rangle\ ,
\label{a2}
\end{align}
where $k$ includes the $N=N^{}_L+N^{}_R$ states $\{\bar l\}$, $\{\bar r\}$.  The time-dependent Schr\"{o}dinger equation for this wave function (setting $\hbar=1$) is
\begin{align}
i\partial_t^{}|\Psi(t)\rangle =H(t)|\Psi(t)\rangle\ .
\label{a5}
\end{align}

In order to solve Eq.~(\ref{a5}) we use an {\it Ansatz} for the total wave function $|\Psi(t)\rangle$, by taking it as a (Slater) product of single-electron wave functions,
\begin{align}
|\Psi(t)\rangle =\prod_k\hat\Phi_{}^{(k)\dagger}(t)|0\rangle\ ,
\label{a3}
\end{align}
where now we have $k=0,~1,~2,\dots~,N$, and
\begin{align}
\hat\Phi_{}^{(k)\dagger}(t)=\sum_{l}b_{l}^{(k)}(t)a_{l}^\dagger +b_{0}^{(k)}(t)a_0^\dagger+\sum_{r}b_{r}^{(k)}(t)\,a_{r}^\dagger\ ,
\label{a4}
\end{align}
with the initial conditions\cite{com1}
\begin{align}
b_{l}^{(k)}(0)=\delta^{}_{kl}\delta^{}_{{\bar l}l},\ \  b_{r}^{(k)}(0)=\delta^{}_{kr}\delta^{}_{{\bar r}r},\ \   b_{0}^{(k)}(0)=n^{}_0\delta^{}_{k0}.
\label{incond}
\end{align}

For convenience, we now order the $N$ states in the products (\ref{a2}) and ({\ref{a3}) by an increasing index, $k=0,~1,~2,~\dots,~N$.
Substituting Eqs.~(\ref{a3}) into Eq.~(\ref{a5}), the latter can be rewritten as
\begin{align}
&
\sum_k\prod_{k'<k}\hat\Phi_{}^{(k')\dagger}(t)\,\big[i\,
\partial_t\hat\Phi_{}^{(k)\dagger}(t)\big]
\prod_{k''>k}\hat\Phi_{}^{(k'')\dagger}(t)|0\rangle
\nonumber\\
&=
\sum_k\prod_{k'<k}\hat\Phi_{}^{(k')\dagger}(t)\,\Big[
H(t),\hat\Phi_{}^{(k)\dagger}(t)\Big]
\prod_{k''>k}\hat\Phi_{}^{(k'')\dagger}(t)|0\rangle\ .
\label{app4}
\end{align}
Using Eq.~(\ref{a4}), the commutator reads
\begin{align}
&\Big[H(t),\hat\Phi^{(k)\dagger}(t)\Big]=\sum_{l}
\big[E^{}_{l}b_{l}^{(k)}(t)
+\Omega^{}_Lb_{0}^{(k)}(t)\big]
\,a_{l}^\dagger
\nonumber\\
&+\big[E^{}_0(t)b_{0}^{(k)}(t)+\sum_l\Omega^{}_L\,b_{l}^{(k)}(t)
+\sum_r\Omega^{}_R\,b_{r}^{(k)}(t)
\big]\,a_0^\dagger\nonumber\\[5pt]
&+\sum_{r}\big[E^{}_rb_{r}^{(k)}(t)+\Omega^{}_Rb_{0}^{(k)}(t)\big]
\,a_{r}^\dagger\ .
\label{app4p}
\end{align}
Equation~(\ref{app4}) shows that a solution of the many-body Schr\"odinger equation~(\ref{a5}) can  indeed be represented as the product~(\ref{a3}), with $\hat\Phi_{}^{(k)\dagger}(t)$ obtained from the equation
\begin{align}
i\, \partial_t\hat\Phi_{}^{(k)\dagger}(t)=\Big[
H(t),\hat\Phi_{}^{(k)\dagger}(t)\Big]\ .
\label{app5}
\end{align}
Substituting Eq.~(\ref{a4}), this equation  can be rewritten explicitly as
\begin{subequations}
\label{a6}
\begin{align}
i\dot {b}_{l}^{(k)}(t)&=E^{}_lb_{l}^{(k)}(t)+\Omega^{}_L\,b_{0}^{(k)}(t)\ ,
\label{a6a}\\
i\dot {b}_{0}^{(k)}(t)&=E^{}_0(t)\,b_{0}^{(k)}(t)+\sum_l\Omega^{}_L\,b_{l}^{(k)}(t)
+\sum_r\Omega^{}_R\,b_{r}^{(k)}(t)\ ,
\label{a6b}\\
i\dot {b}_{r}^{(k)}(t)&=E^{}_r\,b_{r}^{(k)}(t)+\Omega^{}_R\,b_{0}^{(k)}(t)\ .\label{a6c}
\end{align}
\end{subequations}

It follows from Eqs.~(\ref{a6}) that $\partial_t\langle 0|\hat\Phi_{}^{(k)}(t)\hat\Phi_{}^{(k)\dagger}(t)|0\rangle=0$. Together with the initial conditions (\ref{incond}), this shows probability conservation for the single-electron wave function,
\begin{align}
\sum_l |{b}_{l}^{(k)}(t)|^2+|{b}_{0}^{(k)}(t)|^2+
\sum_r |{b}_{r}^{(k)}(t)|^2=1\ .
\label{probcon}
\end{align}
Note that the amplitudes corresponding to different initial single-electron states $k$ are {\em decoupled} in Eqs.~(\ref{a6}). This is a consequence of the quadratic form of the  Hamiltonian in terms of creation and annihilation operators, Eq.~(\ref{a1}), reflecting non-interacting or mean-field cases.

Equations~(\ref{a6a}) and (\ref{a6c}) for the amplitudes $b_{\alpha}^{(k)}(t)$, where $\alpha=l,r$, can be solved explicitly, thus obtaining
\begin{align}
b_{\alpha}^{(k)}(t)=e^{-iE_{\alpha}t}\Big[\delta^{}_{k\alpha}
-\int\limits_{0}^t
i\,\Omega_{\alpha} b_{0}^{(k)}(t')e^{iE_{\alpha}t'}dt'\Big]\ ,
\label{a12p}
\end{align}
where $\Omega^{}_{l,r}=\Omega^{}_{L,R}$.
Substituting these amplitudes into
Eq.~(\ref{a6b}), this equation becomes
\begin{align}
&i\dot {b}_{0}^{(k)}(t)=E^{}_0(t)b_{0}^{(k)}(t)+
\Omega^{}_k e^{-iE_kt}\nonumber\\
&-i\int\limits_0^tb_0^{(k)}(t')\Big[\Omega_L^2\sum_l
e^{iE_l(t'-t)}+\Omega_R^2 \sum_r
e^{iE_r(t'-t)}\Big]dt'\ .
\label{a6bb}
\end{align}
Note that the sums $\sum_{l,r}$ extend over {\em all} the reservoir states ($E^{}_{l,r}$), without any Pauli principle restrictions.
Equation (\ref{a6bb}) holds also for $k=0$, if we denote $\Omega^{}_0=0$.

The continuum limit is obtained by replacing
$\sum_{l,r}\to\int\varrho^{}_{L,R}dE_{l,r}^{}$, where $\varrho_{L,R}^{}$ are the densities of state in the left and the right leads. We assume that, similarly to the tunneling couplings,  $\Omega_{L,R}^{}$, the densities of state $\varrho_{L,R}^{}$ are also energy independent  (so-called wide band approximation). Using $\int_{-\infty}^\infty e^{iE_{l,r}(t'-t)}dE_{l,r}^{}=2\pi\delta (t'-t)$ and $\int_0^t dt' b_0^{(k)}(t')\delta(t'-t)=b_0^{(k)}(t)/2$,
Eq.~(\ref{a6bb}) becomes
\begin{align}
\frac{d}{dt}{b}_0^{(k)}(t)
=-i\left[E_0(t)-i\Gamma/2\right]b_{0}^{(k)}(t)-i\Omega_ke^{-iE^{}_kt}\
, \label{a6fx}
\end{align}
where we denoted $\Gamma=\Gamma_L^{}+\Gamma_R^{}$, with $\Gamma_{L,R}^{}=2\pi |\Omega_{L,R}^{}|^2\varrho_{L,R}^{}$.

For $k=0$,  Eq. (\ref{a6fx}) becomes
\begin{align}
\frac{d}{dt}b_0^{(0)}(t)
=\Big[-iE_0(t)-(\Gamma/2)\Big]b_{0}^{(0)}(t)\ . \label{b00t}
\end{align}
Solving this equation we find
\begin{align}
b_0^{(0)}(t)=n^{}_0e^{-i\int_0^tE_0(t')dt'-\Gamma t/2}\ .
\label{b001}
\end{align}

For $k\ne 0$ we replace $b_0^{(k)}(t)$ by
\begin{align}
b_0^{(k)}(t)=\widetilde{b}_0^{(k)}(t)e^{-iE_kt}\ ,
\label{b00}
\end{align}
Eq. (\ref{a6fx}) becomes
\begin{align}
\frac{d}{dt}\widetilde{b}_0^{(k)}(t)
=i\left[E_k-E_0(t)+i\Gamma/2\right]\widetilde{b}_{0}^{(k)}(t)-i\Omega_k\
, \label{a6f}
\end{align}
with the solution
\begin{align}
\widetilde{b}_0^{(k)}(t)=-i\,\Omega_k\int\limits_0^t
e^{(iE_k-\Gamma/2)(t-t')
-i\int_{t'}^tE_0(t'')dt''}dt'\ .
\label{b0k}
\end{align}
The amplitudes  $b_{l,r}^{(k)}(t)$ are obtained by substituting Eq.~(\ref{b00}) or Eq. (\ref{b001}) into Eq. (\ref{a12p}). As a result, we obtain the entire wave function $|\Psi (t)\rangle$, Eq.~(\ref{a3}), for any particular initial pure state. This wave function will then be used to calculate the averages of all the charges and currents.

\section{charges and currents}\label{III}

\subsection{Charge on the dot}

For the pure state $|\Psi(t)\rangle$, the average particle numbers (or charges, in units of the electron charge)  inside the quantum dot and the leads, $Q^{}_0(t)$ and $Q^{}_{L,R}(t)$, are given by
\begin{subequations}
\label{charge}
\begin{align}
&Q^{}_{0}(t)=\langle\Psi (t)|a^\dagger_{0}a_{0}^{}
|\Psi (t)\rangle\ ,\label{chargea}\\
&Q^{}_{L(R)}(t)=\Big\langle\Psi (t)\Big|\sum_{l(r)}a^\dagger_{l(r)} a_{l(r)}^{}
\Big|\Psi (t)\Big\rangle\ .
\label{chargeb}
\end{align}
\end{subequations}
Using Eqs.~(\ref{a3}), (\ref{a4}) these quantities can be represented in a very simple form. The charge on the dot $Q^{}_{0}(t)$ can be written as
\begin{align}
Q^{}_0=\langle 0|\hat\Phi_{}^{(N) }\cdots\hat\Phi_{}^{(1)}
a^\dagger_{0}a_{0}^{}\hat\Phi_{}^{(1)\dagger}\cdots\hat\Phi_{}^{(N)\dagger}
|0\rangle\ ,
\end{align}
where we enumerated the initially occupied state in the reservoirs  as $k=\{1,2,\ldots N\}$. Using Eq.~(\ref{a4}) we can write
\begin{align}
&Q^{}_0(t)=
\langle 0|\hat\Phi_{}^{(N)}\cdots\hat\Phi_{}^{(2)}
\hat\Phi_{}^{(2)\dagger}\cdots\hat\Phi_{}^{(N)\dagger}
|0\rangle|b_0^{(1)}(t)|^2\nonumber\\
&+\langle 0|\hat\Phi_{}^{(N) }\cdots\hat\Phi_{}^{(2) } a^\dagger_{0}\hat\Phi_{}^{(1)}
\hat\Phi_{}^{(1)\dagger}a_{0}^{}\Phi_{}^{(2)\dagger }\cdots\hat\Phi_{}^{(N)\dagger}
|0\rangle \nonumber\\
&=|b_0^{(1)}(t)|^2\nonumber\\
&+\langle 0|\hat\Phi_{}^{(N) }\cdots\hat\Phi_{}^{(2) } a^\dagger_{0}\hat\Phi_{}^{(1)}
\hat\Phi_{}^{(1)\dagger}a_{0}^{}\Phi_{}^{(2)\dagger }\cdots\hat\Phi_{}^{(N)\dagger}
|0\rangle\ .
\end{align}
The second step used the identity $\langle 0|\hat\Phi_{}^{(N) }\cdots\hat\Phi_{}^{(2)}
\hat\Phi_{}^{(2)\dagger}\cdots\hat\Phi_{}^{(N)\dagger}
|0\rangle =1$, which is the normalization of the ($N-1$) electron wave function.

Repeating this procedure, we have
\begin{align}
&Q^{}_0(t)=|b_{0}^{(1)}(t)|^2+|b_{0}^{(2)}(t)|^2\nonumber\\
&+\langle 0|\hat\Phi_{}^{(N) }\cdots a^\dagger_{0}\hat\Phi_{}^{(2) } \hat\Phi_{}^{(1)}
\hat\Phi_{}^{(1)\dagger}\Phi_{}^{(2)\dagger }a_{0}^{}\cdots\hat\Phi_{}^{(N)\dagger}
|0\rangle \ .
\end{align}
After $N$ such steps  the last term becomes $\langle 0|a^\dagger_{0}\cdots a^{}_{0}|0\rangle$, which is zero. In the same way we evaluate $Q^{}_{L(R)}(t)$, finally obtaining
\begin{subequations}
\label{charge1}
\begin{align}
&Q^{}_{0}(t)=\sum_k|b_{0}^{(k)}(t)|^2\ ,
\label{charge1a}\\
&Q^{}_{L(R)}(t)
=\sum_k\sum_{l(r)}|b_{l(r)}^{(k)}(t)|^2\ .
\label{charge1b}
\end{align}
\end{subequations}
Here the sum over $l,r$ extends over {\it all} the states (occupied or not) of the leads, whereas the sum over $k$ includes only the $N$ initially
occupied levels. Note that the total charge is conserved in time, so that $Q^{}_L(t)+Q^{}_R(t)+Q^{}_0(t)=N+n_0^{}$. This also follows explicitly from the normalization condition (\ref{probcon}).

Equation (\ref{charge1a}) can be written as
\begin{align}
Q^{}_0(t)\equiv Q^0_0(t)+ Q^L_0(t)+Q^R_0(t)\ ,
\label{QXXX}
\end{align}
where
\begin{align}
&Q^0_0(t)\equiv|b^{(0)}_0(t)|^2=n^{}_0e^{-\Gamma t}\ ,\nonumber\\
&Q^{L,R}_0(t)=\sum_{\bar l,\bar r}|\widetilde b_{0}^{(\bar l,\bar r)}(t)|^2\ .
\label{Q0LR}
\end{align}
The expression for $Q^0_0(t)$ follows from Eq. (\ref{b001}), and $Q^{L,R}_0(t)$ describes the charge arriving to the dot from the left or right lead.
In the continuum limit, this charge is
\begin{align}
Q^{L,R}_0(t)=\varrho^{}_{L,R}\int\limits_{-\infty}^{\mu^{}_{L,R}}
dE^{}_{\bar l,\bar r}|{\widetilde
b}_0^{(\bar l,\bar r)}(t)|^2\ ,
\label{QQQ}
\end{align}
where $\widetilde{b}_0^{(\bar l,\bar r)}(t)$ is given by Eq. (\ref{b0k}).

\subsection{Currents}

 We next discuss the (electron) currents in the left and the right leads. These are given by
\begin{align}
I_{L(R)}(t)&={d\over dt} Q_{L(R)}(t)=\sum_k\sum_{l(r)}{d\over dt}|b_{l(r)}^{(k)}(t)|^2\ ,
\end{align}
where $Q_{L(R)}(t)$ is given by Eq.~(\ref{charge1b}). Splitting the sum over $k$, we have
\begin{subequations}
\label{rescur2}
\begin{align}&I^{}_{R}(t)=I^{}_{0\to R}(t)+ I^{}_{L\to R}(t)+I^{}_{R\to R}(t)\ ,
\label{rescur2a}\\
&I^{}_{L}(t)= I^{}_{0\to L}(t)+I^{}_{L\to L}(t)+I^{}_{R\to L}(t)\ ,
\label{rescur2b}
\end{align}
\end{subequations}
where
\begin{align}
I^{}_{0\to R}(t)={d\over dt}\sum_r|b_r^{(0)}(t)|^2
\end{align}
is the right lead current coming from the quantum dot,
\begin{align}
I^{}_{L\to R}(t)={d\over dt}\sum_{\bar l, r}|b_r^{(\bar l)}(t)|^2
\end{align}
is the right-lead current originated by electrons initially occupied in the left lead and
\begin{align}
I^{}_{R\to R}(t)={d\over dt}\sum_{\bar r, r}|b_r^{(\bar r)}(t)|^2
\end{align}
 is the same for electrons initially occupied in the right lead.
 Similarly $I^{}_{0\to L}$, $I^{}_{R\to L}(t)$ and $I^{}_{L\to L}(t)$ denote the same components of the left-lead current.

Consider first the current from the left to right lead, $I^{}_{L\to R}(t)$. Using Eqs.~(\ref{a6c}) and (\ref{a12p}) we have
\begin{align}
{d\over dt}|b_r^{(\bar l)}(t)|^2&=2{\rm Re}\big[\dot{b}_r^{(\bar l)}(t)b_r^{(\bar l)*}(t)\big]=2\Omega^{}_R{\rm Im}\big[b_0^{(\bar l)}(t)b_r^{(\bar l)*}(t)\big]\nonumber\\
&=2\Omega^{2}_R{\rm Re}\left[\int\limits_0^t dt' [b_0^{(\bar l)}(t)]^*b_0^{(\bar l)}(t')e^{iE^{}_r(t'-t)}\right]\ .
\label{ddbr}
\end{align}
Changing the sum over $r$ to an integral, and performing the integration over $t'$ as in Eqs.~(\ref{a6bb}), (\ref{a6fx}), we find
\begin{align}
&I^{}_{L\to R}(t)= \Gamma^{}_R\sum_{\bar l}|b_0^{(\bar l)}(t)|^2\equiv \Gamma^{}_R Q^L_0(t)\ .
\label{rescur1}
\end{align}
Comparing with Eqs. (\ref{QXXX}) and (\ref{Q0LR}), we see that the sum in Eq. (\ref{rescur1}), denoted by $Q^L_0(t)$, is the part of $Q^{}_0(t)$ which originates from the left lead.

Equation ({\ref{ddbr}) is also valid when ${\bar l}$ is replaced by $0$, and therefore
\begin{align}
I^{}_{0\to R}(t)=\Gamma^{}_RQ^0_0(t)\ .
\end{align}
 Similarly one obtains
 \begin{align}
 I_{R\to L}(t)=\Gamma_L \sum_{\bar r}|b_0^{(\bar r)}(t)|^2\equiv \Gamma^{}_L Q^R_0(t)\ ,
\label{rescur3}
\end{align}
see Eq. (\ref{Q0LR}).

In order to evaluate the last component of the right-lead current, namely $I_{R\to R}(t)$, in the most simple way, we employ the conservation of probability for a single-electron wave function,  Eq.~(\ref{probcon}).
We obtain
\begin{align}
&I^{}_{R\to R}(t)=-{d\over dt}\sum_{\bar r}\Big[\sum_l |b_{l}^{(\bar r)}(t)|^2+|b_{0}^{(\bar r)}(t)|^2\Big]\nonumber\\[5pt]
&\ \ \ =-I_{R\to L}^{}(t)
-\dot Q_0^{R}(t)=-\Gamma^{}_LQ^R_0(t)-\dot Q_0^{R}(t)\ ,
\label{probcon1}
\end{align}
  Altogether, we end up with
\begin{align}
I^{}_R(t)=\Gamma^{}_R\big[Q^L_0(t)+Q^0_0(t)\big]-\Gamma^{}_LQ^R_0-\dot Q^R_0\ .
\label{IRa}
\end{align}
The left hand lead current is given by the same formula, with $R\leftrightarrow L$ everywhere.

\subsection{Time independent dot energy}\label{nonoise}

For a time-independent single level on the dot, at $E^{}_0(t)=\overline E^{}_0=0$,  Eq. (\ref{b0k}) yields
\begin{align}
\widetilde{b}^{(k)}_0(t)=\Omega^{}_k\frac{1-e^{(iE^{}_k-\Gamma/2)t}}{E^{}_k+i\Gamma/2}\ ,
\label{b0no}
\end{align}
 hence
 \begin{align}
 |\widetilde{b}^{(k)}_0(t)|^2 \equiv\frac{2\Omega^2_k}{\Gamma}{\cal B}(E,t)\ ,
 \end{align}
where $E=E^{}_k$ and
\begin{align}
{\cal B}(E,t)=\frac{\Gamma\big[1-2\cos(E t)e^{-\Gamma t/2}+e^{-\Gamma t}]}{2\big(E^2+\Gamma^2/4\big)}\ .
\label{Bno}
\end{align}
Equation (\ref{QQQ}) thus becomes
\begin{align}
Q^{L,R}_0(t)=\frac{\Gamma^{}_{L,R}}{\Gamma}\int_{-\infty}^{\mu^{}_{L,R}}\frac{dE}{\pi}{\cal B}(E,t)\ .
\label{QLRno}
\end{align}

Without bias, $\mu^{}_L=\mu^{}_R=\mu$,  the charge on the dot is [Eqs. (\ref{QXXX}), (\ref{Q0LR}) and (\ref{QLRno})],
\begin{align}
Q^{}_0(\mu,t)=n^{}_0e^{-\Gamma t}+\int_{-\infty}^{\mu}\frac{dE}{\pi}{\cal B}(E,t)\ .
 \end{align}
The argument $\mu$ was added on the left hand side to facilitate further discussions.  In the steady state limit, $t\to\infty$, we find
 \begin{align}
Q^{}_0(\mu,\infty)=1/2+\arctan(2\mu/\Gamma)/\pi\ .
\end{align}
The charge on the dot increases from 0 to 1 as the chemical potential increases from $-\infty$ to $\infty$, see black solid line in Fig. \ref{Qst}. The main increase occurs in the range $|\mu|<\Gamma/2$.

Substituting Eq. (\ref{QLRno}) into Eq. (\ref{IRa}), the  current in the right lead is
\begin{align}
I^{}_R(t)=\frac{\Gamma^{}_L\Gamma^{}_R}{\Gamma}\int_{\mu^{}_R}^{\mu^{}_L}\frac{dE}{\pi}{\cal B}(E,t)+\Gamma^{}_RQ^0_0(t)-\dot{Q}^R_0(t)\ ,
\label{IRno}
\end{align}
where $Q^0_0(t)$ and $Q^R_0(t)$ are given by Eqs.~(\ref{Q0LR}) and (\ref{QLRno}).  In the steady state limit, the last two terms decay to zero, and the first term describes the current generated by the bias voltage,
\begin{align}
 I^{}_R(\infty)=\frac{\Gamma^{}_L\Gamma^{}_R}{\Gamma}\int_{\mu^{}_R}^{\mu^{}_L}\frac{dE}{\pi} \overline{\cal B}(E)
\approx \frac{\Gamma^{}_L\Gamma^{}_R}{\pi\Gamma}\overline{\cal B}(\mu)(\mu^{}_L-\mu^{}_R)\ ,
\label{curss}
\end{align}
where $\overline{\cal B}(E)={\cal B}(E,\infty)$ and where the last step is the linear response result, for a small bias voltage $eV=\mu^{}_L-\mu^{}_R$ and with $\mu=(\mu^{}_L+\mu^{}_R)/2$. At the steady state, this is also equal to the net current through the dot, $I=I^{}_R=-I^{}_L$. Adding the units $e^2/\hbar$, the conductance of spinless particles is given by  the Landauer formula\cite{lan} at zero temperature, ${\cal G}= I^{}_R(\infty)/V=(e^2/h)T$, with the spinless transmission
\begin{align}
T(\mu)=\frac{2\Gamma^{}_L\Gamma^{}_R}{\Gamma}\overline{\cal B}(\mu)\ ,
\label{Tmu}
\end{align}
reproducing the usual Breit-Wigner form, $T(\mu)=\Gamma^{}_L\Gamma^{}_R/[\mu^2+(\Gamma/2)^2]$, as shown by the black solid line in Fig. \ref{Ist}.

Starting at $t=0$ with $Q^{}_0(\mu,0)=n^{}_0$, the time-dependent charge $Q^{}_0(\mu,t)$ approaches this steady state with two terms, one oscillating with frequency of order $\mu$ and with a decay rate $\Gamma/2$ and the other decaying as $e^{-\Gamma t}$, see black solid lines in Fig. \ref{Qt}.
For $\mu\ne 0$, the oscillations in $Q^{}_0(\mu,t)$ result from electron transitions between the dot and the leads, and they reflect quantum interference between these transitions.  As we discuss below, and as seen in the figures, the noise smears these oscillations. Also, there are no oscillations at $\mu=0$ (see below).

\section{telegraph noise}

\subsection{Markov process}

 The noise distribution of the variables $\{\xi(t)\}$ in Eq. (\ref{noise2}) is characterized by the probabilities
 $p_{\xi_0,\xi}^{}(t_0^{},t)$ to find $\xi(t)$ at the value $\xi=\pm 1$ on the condition that $\xi (t_0)=\xi_0^{}$ [or $p_{\xi_0,\xi}^{}(t_0,t_0)=\delta_{\xi_0,\xi}^{}$].
Therefore,
\begin{align}
{d\over dt}p_{\xi_0,1}^{}(t_0^{},t)&=-\frac{\gamma}{
2}[p_{\xi_0,1}^{}(t_0,t)-p_{\xi_0,-1}^{}(t_0,t)]\nonumber\\
&=-\frac{\gamma}{2}\, [2p_{\xi_0,1}^{}(t_0,t)-1]\ ,
\label{ratenoise}
\end{align}
where $\gamma/2$ is the rate of jumps between two value of $\xi(t)$. In Eq.~(\ref{ratenoise}) we have used  $p_{\xi_0,1}^{}(t_0^{},t)+p_{\xi_0,-1}^{}(t_0,t)=1$, from probability conservation.  The solution of this equation is
\begin{align}
p^{}_{\xi_0,\xi}(t_0,t)=(\delta_{\xi_0,\xi}^{}-1/2)e^{-\gamma (t-t_0)}+1/2\ ,
\label{17}
\end{align}
approaching the steady state value $p^{}_{\xi_0,\xi}(t_0,t)\to p^{}_\xi(t)=1/2$ in the steady state limit $t-t_0\to\infty$, independent of the initial condition $(\xi_0)$.

For any given initial value of $\xi(t_0)$,
the physical quantities
(e.g. the charge on the dot) at time $t$ depend on the specific history of the changes in
$\xi(t)$. We denote the average over all these histories
 by
angular brackets, $\langle...\rangle$. For example,
\begin{align}
\langle\xi (t)\rangle=\sum_{\xi=\pm 1} \xi p^{}_{\xi_0,\xi}(t_0^{},t)=(\delta_{\xi_0, 1}-\delta_{\xi_0, -1})e^{-\gamma(t-t_0)}\ .
\label{18}
\end{align}
In the steady state limit ($t_0\to-\infty$) one has $\langle \xi(t)\rangle=0$.

The noise correlator is generally defined as
\begin{align}
&\langle\xi (t_1^{})\xi (t_1+\tau)\rangle=\sum_{\xi_1,\xi=\pm
1}\xi_1\, \xi\,p_{\xi_1}^{}(t_1)\, p_{\xi_1,\xi}^{}(t_1,t_1+\tau)\ ,
\label{excor}
\end{align}
where $t_1$ corresponds to the steady state, $p_{\xi_1}^{}(t_1)=\pm 1/2$.

Substituting Eq.~(\ref{17}) into Eq.~(\ref{excor}), the correlator becomes
\begin{align}
\langle\xi (t_1)\xi
(t_1+\tau)\rangle=e^{-\gamma\tau}\ .
\label{21}
\end{align}
It is customary to characterize the noise by its power spectrum,
\begin{align}
S(\omega )=2{\rm Re}\int\limits_0^\infty \langle\xi (t_0)\xi
(t_0+\tau)\rangle e^{i\omega\tau}d\tau =\frac{2\gamma}{\omega^2+\gamma^2}\ .
\label{spect}
\end{align}

\subsection{Electron motion under the noise}

We now consider the effects on the telegraph noise on the measurable quantities, $Q^{}_0(t)$ and $I^{}_{R,L}(t)$.
To evaluate $\langle Q^{L,R}_0(t)\rangle$ from Eq. (\ref{QQQ}), we need the average $\langle |{\widetilde b}_0^{(k)}(t)|^2\rangle$. It turns out convenient to express this average in terms of  $\langle \widetilde{b}_{0}^{(k)}(t)\rangle$.
Multiplying Eq. (\ref{a6f}) by ${\widetilde b}_0^{(k)*}(t)$, we find
\begin{align}
\frac{d}{dt}&|{\widetilde b}_0^{(k)}(t)|^2=2{\rm Re}\big[{\widetilde
b}_0^{(k)*}(t)\frac{d}{dt}{\widetilde
b}_0^{(k)}(t)\big]\nonumber\\
&=2{\rm Re}\Big[{\widetilde
b}_0^{(k)*}(t)\Big(i[E^{}_k-E^{}_0(t)+i\Gamma/2]{\widetilde
b}_0^{(k)}(t)-i\Omega^{}_k\Big)\Big]\nonumber\\
&=-\Gamma|{\widetilde b}_0^{(k)}(t)|^2-2\Omega^{}_k{\rm
Im}\big[{\widetilde b}_0^{(k)}(t)\big]\ .
\label{diff}
\end{align}
Solving this differential equation and then averaging over the noise we obtain
\begin{align}
\langle |{\widetilde b}_0^{(k)}(t)|^2\rangle=-2\Omega^{}_k \int\limits_0^t
e^{\Gamma (t'-t)}{\rm Im}\Big[\langle {\widetilde
b}_0^{(k)}(t')\rangle\Big] dt'\ . \label{rate3avsa}
\end{align}

From Eq. (\ref{IRa}), the currents are also determined by the partial charges $Q^0_0(t)$ [which does not depend on the noise, Eq. (\ref{QXXX})], and by $\langle Q^{L,R}_0(t)\rangle$, which depend only on $\langle{\widetilde b}_0^{(k)}(t)\rangle$.
Thus, it is sufficient to calculate the latter average.
To obtain this average,
 we average Eq.~(\ref{a6f}) for $k\ne 0$ over the noise,
\begin{align}
\frac{d}{dt}\langle\widetilde{b}_0^{(k)}(t)\rangle &=i[E_k+i\Gamma/2]\langle\widetilde{b}_{0}^{(k)}(t)\rangle\nonumber\\
&-iU\langle\xi (t)\widetilde{b}_{0}^{(k)}(t)\rangle/2-i\Omega_k\ .
\label{a6e}
\end{align}
(Here and in the following we choose a scale where $\overline E_0=0$). To evaluate the term $\langle\xi (t) \widetilde{b}_0^{(k)}(t)\rangle$, we  multiply Eq.~(\ref{a6f}) by $\xi (t)$, taking into account that $\xi^2(t)=1$, and average the resulting equation. The result is
\begin{align}
\langle\xi(t)\frac{d}{dt} \widetilde{b}_0^{(k)}(t)\rangle =&i[E_k+i\Gamma/2]
\langle\xi (t)  \widetilde{b}_{0}^{(k)}(t)\rangle\nonumber\\
&~~~~~~~~-iU\langle \widetilde{b}_{0}^{(k)}(t)\rangle/2\ .
\label{a7d}
\end{align}

In the case of an exponential noise-correlator, Eq.~(\ref{21}), one can derive the following ``differential formula''  \cite{shapiro,klya}, which is extremely useful for treating stochastic equations,
\begin{align}
\langle\xi(t){d\over dt}\widetilde{b}_0^{(k)}(t)\rangle&={d\over dt} \langle\xi(t) \widetilde{b}_0^{(k)}(t)\rangle
+\gamma \langle\xi (t)\widetilde{b}_0^{(k)}(t)\rangle\ ,
\label{df}
\end{align}
where $\gamma$ is the width of the noise-spectrum, Eq.~(\ref{spect}).
A brief discussion of this formula, especially in the present context, is given in Appendix \ref{app1}.

Using Eq.~(\ref{df}),  Eq.~(\ref{a7d}) becomes
\begin{align}
{d\over dt}\langle\xi(t) \widetilde{b}_0^{(k)}(t)\rangle =&i[E_k+i(\Gamma +2\gamma)/2]
\langle\xi (t) \widetilde{b}_{0}^{(k)}(t)\rangle\nonumber\\
&~~~~~~~~-iU\langle \widetilde{b}_{0}^{(k)}(t)\rangle/2\ ,
\label{a6cc}
\end{align}
and we end up with two coupled differential equations, (\ref{a6e}) and (\ref{a6cc}), in the two functions $\langle \widetilde{b}_{0}^{(k)}(t)\rangle$ and $\langle \widetilde\xi {b}_{0}^{(k)}(t)\rangle$.
Below we describe the solutions of these equations. Having these solutions, the calculations of the average charge and currents [Eqs. (\ref{Q0LR}), (\ref{IRa})] are  carried out using Eq. (\ref{rate3avsa}).

\section{The steady state}

In the steady state limit, $t\to\infty$, the left hand sides of Eqs. (\ref{a6e}), (\ref{a6cc}) vanish, and become algebraic:
\begin{align}
&\Big(E_k+i{\Gamma\over2}\Big)\langle\widetilde{b}_{0}^{(k)}(\infty)\rangle
-{U\over2}\langle\xi \widetilde{b}_{0}^{(k)}(\infty)\rangle=\Omega_k\ ,
\label{a6esta}\\[5pt]
&\Big(E_k+i{\Gamma+2\gamma\over2}\Big)
\langle\xi \widetilde{b}_{0}^{(k)}(\infty)\rangle-{U\over2}\langle \widetilde{b}_{0}^{(k)}(\infty)\rangle =0\ .
\label{a6estb}
\end{align}
Substituting $\langle\xi \widetilde{b}_{0}^{(k)}(\infty)\rangle$ from Eq.~(\ref{a6estb}) into Eq.~(\ref{a6esta}), the latter yields $\langle\widetilde{b}_{0}^{(k)}(\infty)\rangle\equiv \Omega^{}_k\overline{b}(E^{}_k)$, with
\begin{align}
\overline{b}(E)\equiv\frac{1}{ E[1-\eta(E)]+i\big[\Gamma+\eta(E)(\Gamma+2\gamma)\big]/2}\  ,
\label{b000}
\end{align}
where
\begin{align}
\eta(E)=\frac{U^2}{4E^2+\big(\Gamma+2\gamma\big)^2}\ .
\label{a66e}
\end{align}

Comparing Eq.~(\ref{b000}) with the $t\to \infty$ limit of Eq.~(\ref{b0no}),
it looks as if the noise introduces a complex energy-dependent shift in $(E-\overline{E}^{}_0)$, the difference between the electron's energy and the energy on the dot without the noise. One might be tempted to  introduce this shift into observable quantities, and proceed as in the case without noise, Sec. \ref{nonoise}. However, this is not necessarily justified\cite{complex}:  in the present case, the physical measurable quantities are given by $\langle|\widetilde{b}_{0}^{(k)}(\infty)|^2\rangle$, and {\it not} by
$|\langle\widetilde{b}_{0}^{(k)}(\infty)\rangle|^2$!

As we show below, the former average is proportional to ${\rm Im}[\overline{b}(E)]$. The real part of the shift in $E$ leads to a renormalization of $(E-\overline{E}^{}_0)$ by a factor $[1-\eta(E)]$. This appears to be a most noticeable effect of the noise: it can greatly increase the current for  $\eta(E)\sim 1$. The imaginary part of the shift renormalizes the parameter $\Gamma$, yielding  faster relaxation rates and broader conductance peaks (see below).

\subsection{Charge on the dot}

The calculation of the charge on the dot [Eqs.~(\ref{Q0LR}), (\ref{QQQ})] requires the average $\langle|\widetilde{b}_{0}^{(k)}(\infty)|^2\rangle$.  To obtain this average, one should start by setting the LHS of Eq.~(\ref{diff}) to zero, and then average, ending up with
\begin{align}
\langle|\widetilde{b}^{(k)}_0(\infty)|^2\rangle=-\frac{2\Omega^{}_k}{\Gamma}{\rm Im}\langle\widetilde{b}_{0}^{(k)}(\infty)\rangle\equiv \frac{2\Omega_k^2}{\Gamma}\overline{\cal B}(E^{}_k)\ ,
\label{Bbar}
\end{align}
where  $\overline{\cal B}(E)\equiv -{\rm Im}\overline{b}(E)$. This equation is reminiscent of the optical theorem in scattering theory.
Therefore [Eq. (\ref{QQQ})]
\begin{align}
\langle Q^{L,R}_0(\infty)\rangle=\frac{\Gamma^{}_{L,R}}{\Gamma}\int_{-\infty}^{\mu^{}_{L,R}}\frac{dE}{\pi}\overline{\cal B}(E)\ .
\label{QstLR}
\end{align}
In this limit, $Q^0_0(\infty)=0$ [Eq. (\ref{Q0LR})], and the total average charge on the dot is given by Eq. (\ref{QXXX}). Without bias, $\mu^{}_L=\mu^{}_R=\mu$, this becomes 
\begin{align}
\langle Q^{}_0(\mu,\infty)\rangle=\int_{-\infty}^{\mu}\frac{dE}{\pi}\overline{\cal B}(E)\ .
\end{align}

In the limit of no-noise, $U=0$, the steady state charge and currents were discussed in Sec. \ref{nonoise}.
With the noise, we split the results into two regions. First, for $\gamma^2>U^2$ we find
\begin{align}
\overline{\cal B}=\frac{A^{}_-}{E^2+\Gamma_-^2/4}-\frac{A^{}_+}{E^2+\Gamma_+^2/4}\ ,
\label{lrggam}
\end{align}
where
\begin{align}
\Gamma^{}_\pm=\gamma+\Gamma\pm\sqrt{\gamma^2-U^2}
\label{Gpm}
\end{align}
 and
\begin{align}
A^{}_\pm=\frac{\gamma\Gamma+U^2\mp\Gamma\sqrt{\gamma^2-U^2}}{4\sqrt{\gamma^2-U^2}}\ .
\end{align}
In this case,  $\overline{\cal B}$ has a single peak around $E=0$, which is slightly broader than that without the noise. The resulting function $Q_0(\mu,\infty)$ looks similar to the one without the noise, see e.g. the dashed green line in Fig. \ref{Qst}. Although the noise now generates two levels on the dot ($E=\pm U/2$), which become randomly occupied, the width of the noise distribution $\gamma$ is larger than the distance between them $U$, resulting in a} single peak. It also follows from  Eq.~(\ref{lrggam}) that in the limit $\gamma\to\infty$ the effect of noise disappears.

Second, for $\gamma^2<U^2$ one can also write
\begin{align}
\overline{\cal B}&=\frac{C^{}_+}{(E+\sqrt{U^2-\gamma^2}/2)^2+(\Gamma+\gamma)^2/4}\nonumber\\
&+\frac{C^{}_-}{(E-\sqrt{U^2-\gamma^2}/2)^2+(\Gamma+\gamma)^2/4}\ ,
\label{smgam}
\end{align}
with
\begin{align}
C^{}_\pm=\Gamma\big(\Gamma+2\gamma\pm 2\gamma E/\sqrt{U^2-\gamma^2}\big)/4\ .
\end{align}
As $U$ increases above $\gamma$, $\overline{\cal B}$ develops two peaks, at $E\approx\pm \sqrt{U^2-\gamma^2}/2$, with the same width, $(\Gamma+\gamma)$, corresponding to the energy levels $E=\pm U/2$, generated by the noise, Eq.~(\ref{noise2}).
As $\mu$ increases from $-\infty$, the electrons first fill mainly the level $E=-U/2$ on the dot. Similarly, when $\mu>0$, the electrons fill mainly the level $E=U/2$.  $Q^{}_0$ reaches  maximum slopes near these peaks, and is almost flat between them. The appearance of this approximate plateau reflects the existence of the two levels.


 From Eq.~(\ref{b000}), one can see that $\overline{b}(E)^*=-\overline{b}(-E)$, and therefore $\overline{B}(E)=\overline{B}(-E)$ is an even function of $E$. Therefore, $\int_{-\infty}^{-\mu}\frac{dE}{\pi}\overline{\cal B}(E)=\int_{\mu}^\infty\frac{dE}{\pi}\overline{\cal B}(E)$. This relation yields
$\langle Q_0(\mu,\infty)\rangle +\langle Q^{}_0(-\mu,\infty)\rangle=\Gamma\Big(\int_{-\infty}^{\mu}\frac{dE}{\pi}\overline{\cal B}(E)+\int_{\mu}^\infty\frac{dE}{\pi}\overline{\cal B}(E)\Big)=Q^{}_0(\infty,\infty)=1$, i.e.
\begin{align}
\langle Q^{}_0(\mu,\infty)\rangle=1-\langle Q^{}_0(-\mu,\infty)\rangle\ ,
\label{sym}
\end{align}
and therefore also $\langle Q^{}_0(0,\infty)\rangle=1/2$. Indeed, all the graphs in Fig. \ref{Qst} obey this symmetry.
  Interestingly, increasing $|U|$ at fixed $\mu$ causes an increase of the the charge on the dot for $\mu<0$, when the level $-U/2$ in filled, and decreases this charge for $\mu>0$, when the level $U/2$ is filled (instead of the ``original" level at $\overline{E}^{}_0=0$).

\begin{figure}[tbh]
\includegraphics[width=7.5cm]{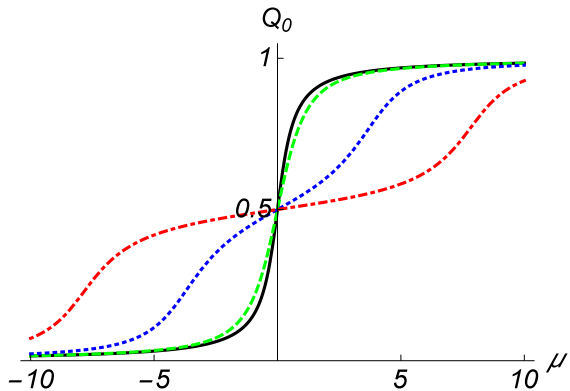}
\caption{(Color online) Average steady-state zero-bias charge on the dot $\langle Q^{}_0\rangle$, when the energy level is fluctuating by the telegraph noise with the rate (spectral width) $\gamma=2$, with amplitude $U=0$ (solid black), $U=1.8$ (dashed green), $U=8$ (dotted blue) and $U=16$ (dot-dashed red). All energies are in units of $\Gamma$.}
\label{Qst}
\end{figure}

\subsection{Steady state currents}

Consider now the steady state current in the right lead. Substituting Eq. (\ref{QstLR}) into Eq. (\ref{IRa}), we find that it again reproduces Eq. (\ref{curss}), where now the LHS is $\langle I^{}_R(\infty)\rangle$ and $\overline{\cal B}$ is defined in Eq. (\ref{Bbar}).  The conductance is again described by the Landauer formula, and the average transmission is given by Eq. (\ref{Tmu}). As mentioned above, $\overline{\cal B}(E)$ -- and therefore also $T(\mu)$ -- are even functions of $\mu$. As seen in Fig.~\ref{Ist}, this function exhibits a single peak at $\mu=0$ for $U^2<\gamma^2$, and a double peak for $U^2>\gamma^2$. These results for the steady state current coincide with those obtained earlier by using very different approach\cite{YG}.
The physical origin of the different behavior in the two limits is easy to understand. For $|U|\ll \gamma$, the fluctuator changes states very quickly, and therefore the properties of the system self-average over these fluctuations, ending up with a single peak in the transmission.  For $|U|\gg \gamma$, the fluctuator stays in each state for a long time before flipping, and then the quantum dot exhibits beats between the two states, ending up with the two peaks in the transmission. Similar two limits also appear for a single qubit coupled to a single fluctuator, see Eq. (35) in Ref. \onlinecite{13}.

\begin{figure}[tbh]
\includegraphics[width=7.5cm]{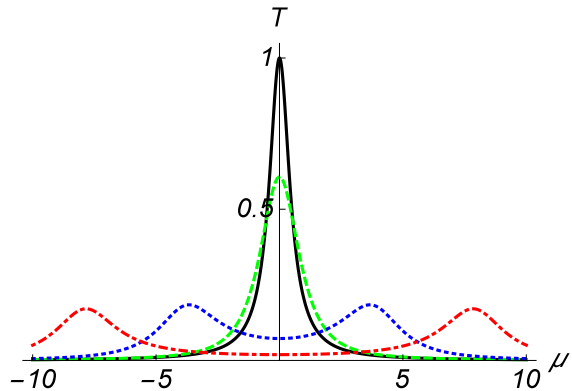}
\caption{(Color online) Average linear response steady-state transmission $T(\mu)$, for the symmetric case $\Gamma^{}_L=\Gamma^{}_R=\Gamma/2$, when the energy level is fluctuating by the telegraph noise with the rate (spectral width) $\gamma=2$, with amplitude $U=0$ (solid black), $U=1.8$ (dashed green), $U=8$ (dotted blue) and $U=16$ (dot-dashed red). All energies are in units of $\Gamma$.}
\label{Ist}
\end{figure}

When $\gamma^2\gg U^2$ the peak in $T(\mu)$ remains at the average energy $\overline{E}^{}_0$, with a smaller height and a slightly larger width.  the dot energy ``jumps" between the two levels very quickly, so that one only observes the average energy.
In contrast,
when $U,~\Gamma\gg\gamma$, Eq. (\ref{smgam}) becomes
\begin{align}
\overline{\cal B}\approx \frac{\Gamma}{4}\Big[\frac{1}{(E-U/2)^2+\Gamma^2/4}
+\frac{1}{(E+U/2)^2+\Gamma^2/4}\Big]\ ;
\label{smallgam}
\end{align}
the transmission is an uncorrelated average of two transmissions, corresponding to the two levels.
This can be easily understood qualitatively: a small hopping rate $\gamma$ means that the dot remains at each of the levels for a relatively long time, which yields the resonant scattering from that level only.
It is interesting to compare this result with the coherent scattering through a dot with two levels, $E^{}_\pm=\pm U/2$, which are connected to each lead\cite{BIH}.  For a coherent scattering through such a dot, the transmission also exhibits two peaks, at $E=\pm U/2$. However, unlike Eq. (\ref{smallgam}), which always gives a finite transmission at $E=0$, this coherent transmission vanishes at $E=0$, reflecting the Fano effect which results from quantum interference. In some sense, the telegraph noise eliminates all these interference effects. Interestingly, when  unitarity is broken by allowing leakage of the electrons out of the system\cite{BIH}, then the transmission through the dot becomes similar to our results for $|U|\gg\gamma$. In both cases, the perturbation causes decoherence.

\vspace{1cm}

\section{Transients}

The solution of  the two coupled differential equations, (\ref{a6e}) and (\ref{a6cc}) is
$\langle \tilde{b}_{0}^{(k)}(t)\rangle \equiv\Omega^{}_kb(E,t)$, with
\begin{align}
&b(E,t) =\frac{1}{D(E)}\Big\{E+i(\Gamma/2+\gamma)
+ie^{[iE-(\Gamma+\gamma)/2]t}\nonumber\\
&\times\Big[{\sinh\Big[{1\over2}\sqrt{\gamma^2-U^2}t\Big]
\over\sqrt{\gamma^2-U^2}}\Big(U^2/2-\gamma(\gamma+\Gamma/2)
+i\gamma E\Big)\nonumber\\
&-\cosh\Big[{1\over2}\sqrt{\gamma^2-U^2}t\Big] (\gamma+\Gamma/2 -i
E)\Big]\Big\}\equiv\Omega^{}_k b(E,t)\ , \label{b01}
\end{align}
where
\begin{align}
D(E)=[E+i(\Gamma+\gamma)/2]^2+(\gamma^2-U^2)/4\ .
\label{DE}
\end{align}
One can split Eq. (\ref{b01}) into two terms,
\begin{align}
b(E,t)=\overline{b}(E)+\delta b(E,t)\ ,
 \label{delb}
 \end{align}
where $\overline{b}(E)$ is the steady state value given by Eq. (\ref{b000}),
while $\delta b(E,t)$ contains two oscillating decaying terms, which for $U=0$ generate the single oscillating term in Eq.~(\ref{b0no}), reflecting quantum interference due to electron transitions between the dot and the leads.
For $\gamma^2>U^2$, the latter terms have two relaxation 
rates, $\Gamma^{}_\pm/2$ [Eq. (\ref{Gpm})] and one oscillation frequency, $E$, which is the difference between the electron's energy in the lead and the average energy ($\overline{E}^{}_0$) on the dot. For $U^2>\gamma^2$, these terms have a single relaxation rate, $(\Gamma+\gamma)/2$, and two oscillation frequencies, $E\pm\sqrt{\gamma^2-U^2}/2$.
 In both cases, the new relaxation rates are larger than $\Gamma/2$, which is the relaxation rate, generated by coupling with the reservoirs, in the absence of the noise [Eq. (\ref{b0no})]. The noise increases the rate of approach to the steady state.

At long times, the rate of approaching the steady state is dominated by the slowest decaying term, with the smallest relaxation rate. This rate depends on $U$. For $|U|<\gamma$, this rate is given by $\Gamma^{}_-/2$. Writing $\Gamma_-^{}=\Gamma+\gamma_d$, the second term can be attributed to decoherence generated by the telegraph noise. For  $|U|\ll\gamma$ one has
\begin{align}
\gamma_d=\gamma-\sqrt{\gamma^2-U^2}\simeq {U^2\over2\gamma}=\left({U\over2}\right)^2S^{}_0(0)\ ,
\label{decoher}
\end{align}
where $S^{}_0(0)$ is the noise power spectrum at zero frequency, Eq.~(\ref{spect}). This relation between the decoherence rate and the zero-frequency noise spectrum coincides with the well-known expression for the inverse ``dephasing'' time, obtained in many models in a weak damping limit \cite{amnon,14,slich,legg}. The self averaging at $|U|\ll \gamma$ is also the source of the motional narrowing of spectral lines in spin resonance experiments \cite{PWA}.

Consider now the temporal behavior of the charge on the dot and the currents in the leads.
Substitution Eq.~(\ref{b01}) into Eqs. (\ref{rate3avsa})  reproduces Eq. (\ref{QLRno}) for $\langle Q^{L,R}_0(t)\rangle$,
with
\begin{align}
{\cal B}(E,t)&=-\Gamma\int_0^t e^{\Gamma(t'-t)}{\rm Im}[b(E,t')]dt'\nonumber\\
&\equiv \overline{\cal B}(E)(1-e^{-\Gamma t})+\delta{\cal B}(E,t)\ ,
\label{delBB}
\end{align}
where $\overline{\cal B}(E)$ is the steady state value, discussed in the previous section, and where $\delta{\cal B}(E,t)$ contains two oscillating decaying terms.

Adding $Q^0_0$ from Eq. (\ref{Q0LR}), the average of Eq. (\ref{QXXX}) becomes
\begin{align}
\langle Q^{}_0\rangle=n^{}_0e^{-\Gamma t}+\langle Q^L_0(t)\rangle+\langle Q^R_0(t)\rangle\ .
\label{Qtot}
\end{align}
Examples of the time evolution of $\langle Q^{}_0(\mu,t)\rangle$ for  $n^{}_0=0$, with $\Gamma_L=\Gamma_R=\Gamma/2$ and $\mu_L=\mu_R=\mu$, are shown in Fig. \ref{Qt}. The case $n^{}_0=1$ is trivially obtained by adding $n^{}_0e^{-\Gamma t}$. At long times, all the graphs approach the steady state values given in Fig. \ref{Qst}. Again, the graphs for $\gamma^2>U^2$ are almost the same as those without noise, e.g. the dashed green lines in the figure.

Similar to the discussion before Eq. (\ref{sym}), Eq. (\ref{b01}) obeys $b(E,t)=-b(-E,t)^*$, hence ${\cal B}(E,t)={\cal B}(-E,t)$. Therefore, $\int_{-\infty}^{-\mu}\frac{dE}{\pi}{\cal B}(E,t)=\int_{\mu}^\infty\frac{dE}{\pi}{\cal B}(E,t)$ and $\int_{-\infty}^{-\mu}\frac{dE}{\pi}{\cal B}(E,t)+\int_{-\infty}^{\mu}\frac{dE}{\pi}{\cal B}(E,t)=\int_{-\infty}^{\infty}\frac{dE}{\pi}{\cal B}(E,t)$. In particular,
$2\int_{-\infty}^{0}\frac{dE}{\pi}{\cal B}(E,t)=\int_{-\infty}^{\infty}\frac{dE}{\pi}{\cal B}(E,t)$, yielding
\begin{align}
\langle Q^{}_0(\mu,t)\rangle+\langle Q^{}_0(-\mu,t)\rangle =2\langle Q^{}_0(0,t)\rangle\ .
\label{sym1}
\end{align}
This relation, which generalizes Eq. (\ref{sym}), is indeed confirmed by the graphs in Fig. \ref{Qt}.

The case $\mu=0$ is particularly interesting: it exhibits a single exponential behavior, with no oscillation, and is completely independent of the noise.  To explain this, look at Eq. (\ref{delb}). The time-dependent function $\delta b(E,t)$ consists of two oscillating terms, whose energy dependence is of the form $f(E,t)e^{iE t}/D(E)$. The integral $\int^\infty_{-\infty}dEf(E,t)e^{i E t}/D(E)$ can be calculated by closing a path in the upper half of the complex $E-$plane. The poles of the denominator are at $-i\Gamma^{}_\pm/2$ for $\gamma^2>U^2$ or at $-i[\Gamma+\gamma\pm\sqrt{U^2-\gamma^2}]/2$ for $U^2>\gamma^2$. In both cases, these poles are in the lower half of the complex $E-$plane, hence this integral vanishes, and therefore also $\int_{-\infty}^\infty dE[\delta{\cal B}(E,t)]=0$. Only the first term in Eq.~(\ref{delBB}) survives, so that  $\langle Q^{}_0(0,t)\rangle=\langle Q^{}_0(0,\infty)\rangle(1-e^{-\Gamma t})+n^{}_0e^{-\Gamma t}$, {\it independent of the noise}. If one wishes to avoid the effects of the noise, one should adjust to common chemical potential to be at the ``original" energy level on the dot, $\mu=\overline{E}^{}_0$. Looking at graphs with $\mu\ne 0$, one sees that the oscillations decrease as $|U|$ increases; the noise smears the oscillations in the time evolution.

Substituting Eq. (\ref{Qtot}) into Eq. (\ref{IRa}), the  current in the right lead is again given by Eq. (\ref{IRno}).
In the steady state limit one has ${\cal B}(E,\infty)=\overline{\cal B}(E)$, $Q^0_0(\infty)=0$, $\dot{Q}^R_0(\infty)=0$, and we also reproduce Eq. (\ref{curss}). The first term in Eq. (\ref{IRno}),
sometimes called  the ``particle'' current \cite{bb}, vanishes at zero bias. It also vanishes if the quantum dot is connected only to one lead ($\Gamma^{}_L=0$ or $\Gamma^{}_R=0$). In these cases, one only has a transient current, which vanishes at $t\to\infty$. This transient current, sometimes called the ``displacement" current \cite{you,bb}, is given by
\begin{align}
&\langle I^{}_{R,D}(t)\rangle=\Gamma^{}_R\langle Q^0_0(t)\rangle-\langle \dot{Q}^R_0(t)\rangle\nonumber\\
&=\Gamma^{}_R\langle Q^0_0(t)\rangle+\Gamma \langle Q^R_0(t)\rangle+\Gamma^{}_R\int_{-\infty}^{\mu^{}_R}\frac{dE}{\pi}{\rm Im}[b(E,t)]\ ,
\label{curt}
\end{align}
where the second step follows from Eq. (\ref{diff}).

\begin{widetext}

\begin{figure}[ht]
\begin{center}

\includegraphics[width=0.3\textwidth,height=0.14\textheight]{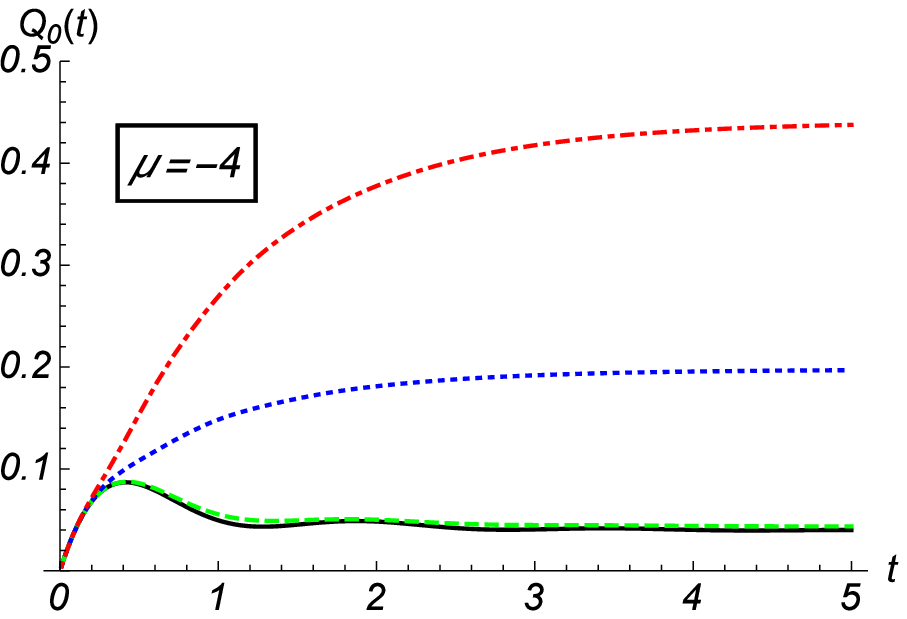}\ \ \
\includegraphics[width=0.3\textwidth,height=0.14\textheight]{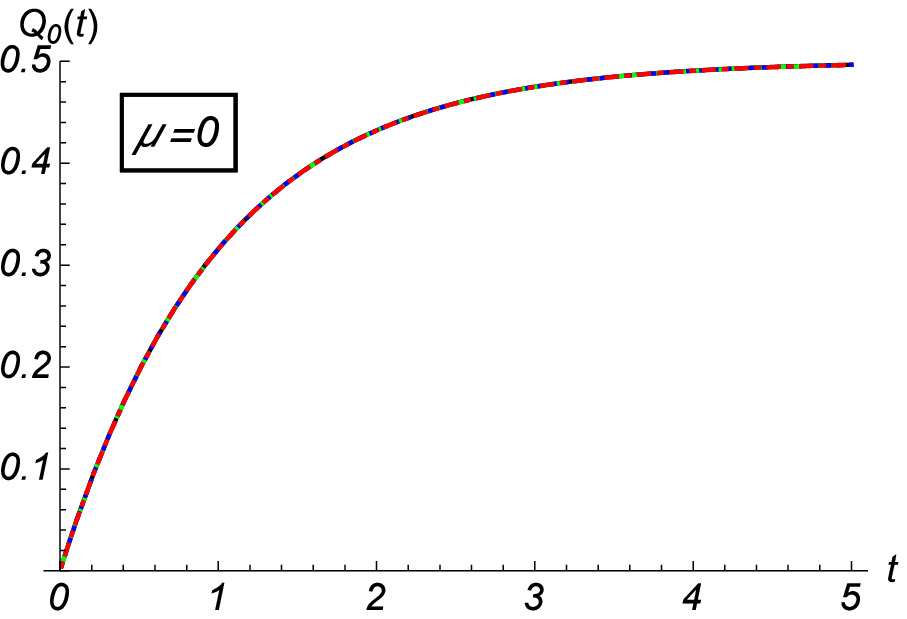}\ \ \
\includegraphics[width=0.3\textwidth,height=0.14\textheight]{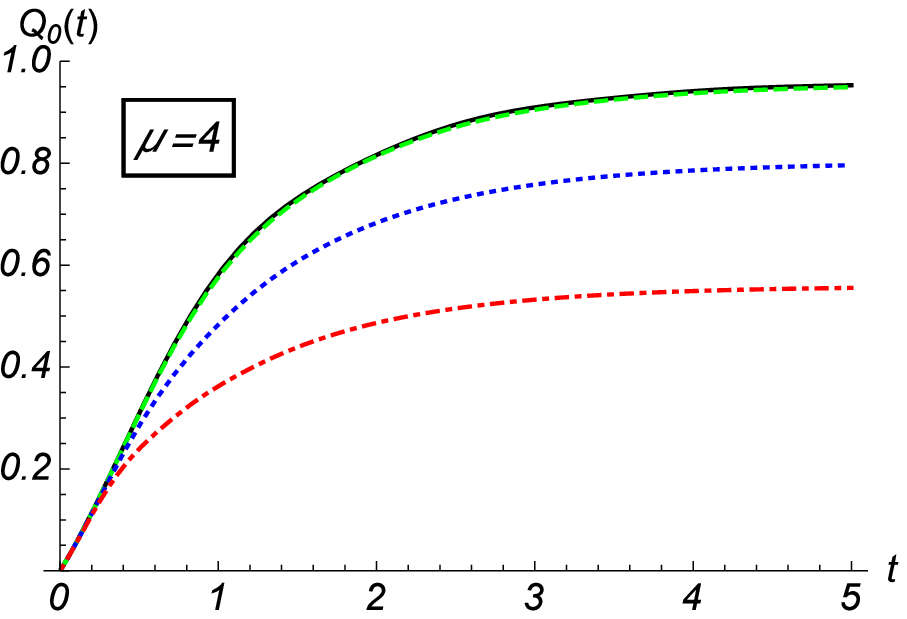}\\
\end{center}
\caption{(Color online) Average charge inside the well as a function of time (in units of $\Gamma^{-1}$) for $\gamma=2$,  $n^{}_0=0$ and different $\mu$'s, and with amplitudes $U=0$ (solid black), $U=1.8$ (green dashed), $U=8$ (dotted blue) and $U=16$ (dot-dashed red).  
All energies are in units of $\Gamma$.}
\label{Qt}
\end{figure}

\begin{figure}[ht]
\begin{center}

\includegraphics[width=0.32\textwidth,height=0.14\textheight]{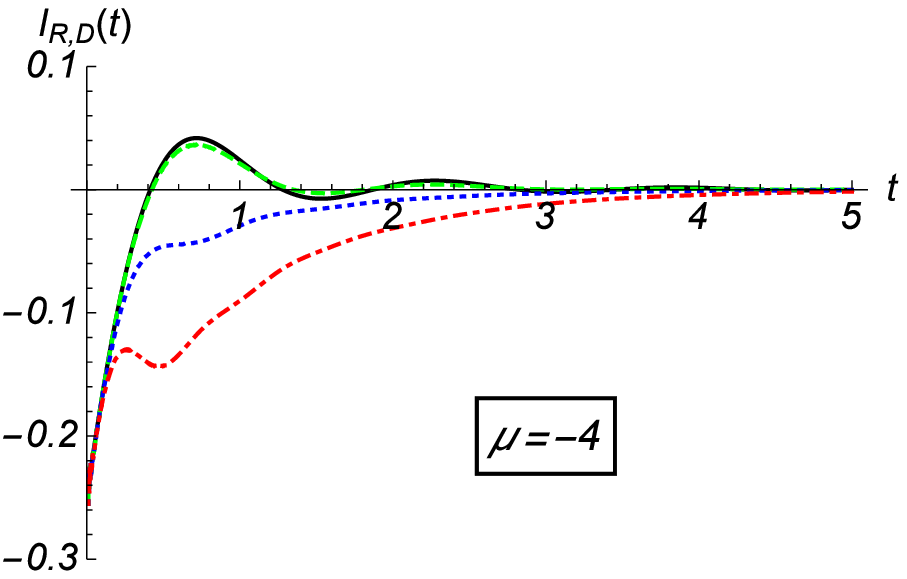}\ \ \
\includegraphics[width=0.3\textwidth,height=0.14\textheight]{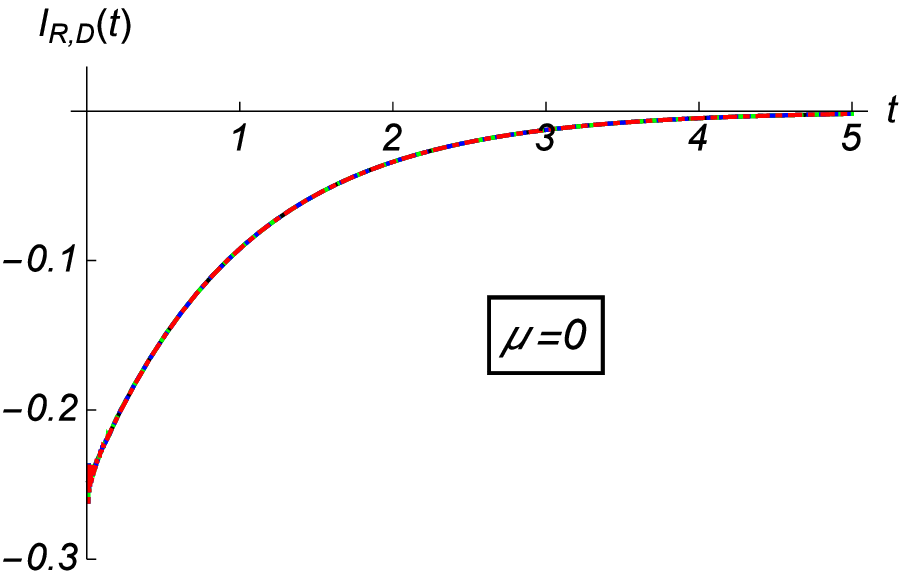}\ \ \
\includegraphics[width=0.3\textwidth,height=0.14\textheight]{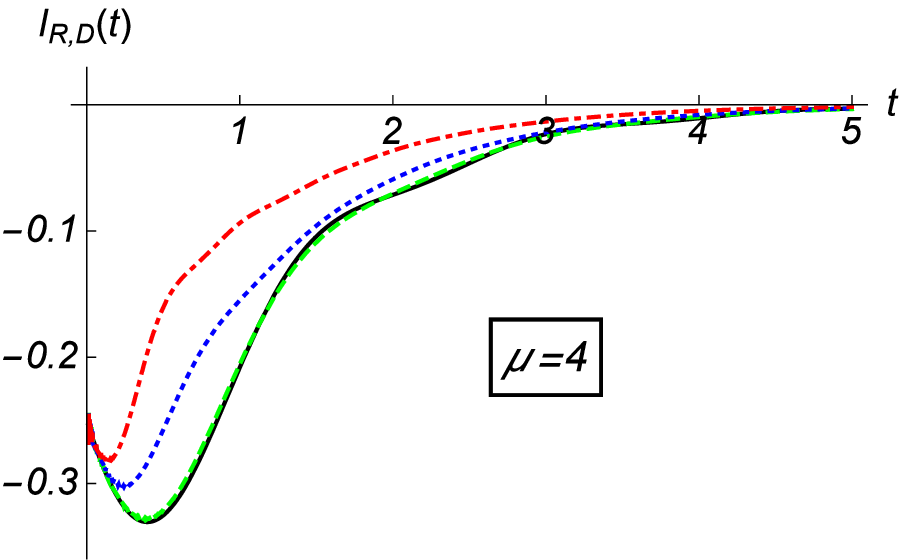}
\end{center}
\caption{(Color online) Displacement current as a function of time (in units of $\Gamma^{-1}$) for $\gamma=2$,  $n^{}_0=0$ and different $\mu$'s, and with amplitudes $U=0$ (solid black), $U=1.8$ (green dashed), $U=8$ (dotted blue) and $U=16$ (dot-dashed red). All energies are in units of $\Gamma$.}
\label{IDDt}
\end{figure}

\end{widetext}

Figure~\ref{IDDt} shows the transient displacement current $I^{}_{R,D}$ for a symmetric case, $\Gamma_L=\Gamma_R=\Gamma/2$, for the same parameters as in Fig. \ref{Qt}. Again, the results for $n^{}_0=1$ are obtained by adding the first term in Eq. (\ref{curt}). Again, the oscillating terms are completely absent for $\mu=0$, and they are smeared out as $|U|$ increases.
Figure \ref{curt1} shows an example of the time evolution of current in the left lead for a finite bias. Except for the case $U=-8$, all the graphs are close to those of the displacement current, shown in Fig.~\ref{Qt}, and approach  the steady state current, Eq. (\ref{curss}) and Fig. \ref{Ist}. The graph for $U=-8$ is different, because then the steady state transmission is close to the resonance associated with the energy level $-U/2=-4$, and the `particle' current feels this resonance already at relatively short times.

\begin{figure}[tbh]
\includegraphics[width=7.5cm]{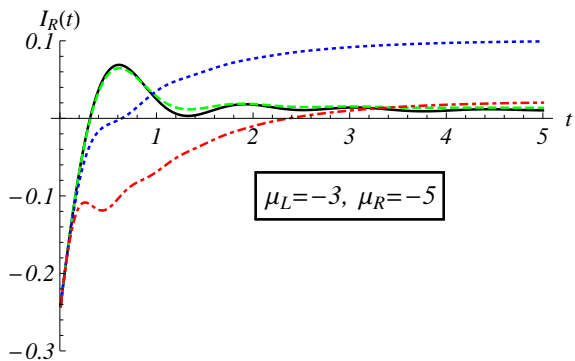}
\caption{(Color online) Average current in the left lead $I^{}_R(t)$, for the same parameters as in Fig. \ref{IDDt}, except for the bias,  $\mu^{}_L=-3$ and $\mu^{}_R=-5$. All energies are in units of $\Gamma$.}
\label{curt1}
\end{figure}

\section{Discussion}

We have demonstrated that the SEA is convenient for averaging over random time-dependent potentials in general, and over telegraph noise in particular. In the present paper we wanted to represent the method and emphasize its advantages for open mesoscopic systems, and therefore we applied it to the simplest case of a single quantum dot, a single fluctuator, one path between the leads, spinless electrons and zero temperature. Needless to say, all of these assumptions can be relieved.

In principle, we could follow Ref. \onlinecite{amnon}, and calculate the conditional probabilities for currents and charges to reach different values at time $t$. Here we preferred to simplify the analysis, and calculate directly the averages over these different values. In the wide band approximation, all the quantities of interest were found to be determined by a single function $b(E,t)$, which is obtained by solving two coupled linear differential equations, and this led to explicit simple expressions for the currents and charges. 

Another possible generalization assumes different jump rates between the two levels\cite{YG,amnon}, or other distributions of the random levels. The formalism can be easily adapted to include such changes.

Even in the simplest case of resonant transmission through a single quantum dot, the telegraph noise generates several interesting results, including the crossover of the effective Landauer transmission from that of a single level to the incoherent sum of the transmissions of the two levels, and the noise dependent effects on the transient time evolution of the charge on the dot and of the currents in the leads. The relaxation rates of the transients turn out to be directly related to the widths of the peaks in the corresponding steady state conductance peaks. The noise broadens the peaks and increases the relaxation rates. The noise also reduces the transient temporary oscillations in the currents and charges. However, when the width of the noise distribution is very large, the effect of the noise on the system disappears. Both the steady state and the transient results exhibit interesting symmetries, Eqs. (\ref{sym}) and (\ref{sym1}).

Usually, the environment is expected to cause decoherence\cite{amnon}. For a single path between the leads one does not have a direct access to the phases of the wave functions, except for the oscillations in the transient behavior of charges and currents, discussed after Eq.~(\ref{delb}). However, the discussion after Eq. (\ref{smallgam}), which compared the noisy results for the steady-state current with the coherent transport through two static levels, does indicate a total loss of coherence.
This issue will become much clearer when we shall study the effects of telegraph noise on transport through an Aharonov-Bohm interferometer, or through a Fano resonance. We plan to address these issues in a future publication.

\begin{acknowledgments}
We thank Yuri Galperin for reminding us of Ref. \onlinecite{YG} and for corresponence related to Ref. \onlinecite{alt1}.  S.G. also thanks Avraham Kofman for useful discussions. This work was supported by the Israeli Science Foundation (Israel), under grant number 252/11, and by the infrastructure program of Israel Ministry of Science and Technology under contract 3-11173.
\end{acknowledgments}

\appendix
\section{The Shapiro-Loginov formula}\label{app1}

Since the Shapiro-Loginov differential formula\cite{shapiro,klya} is  essential for the treatment of noise in the framework of our single-electron approach, this Appendix outlines its derivation \cite{shapiro}, adopted for the specific system studied in the present work. Consider a functional $R[\xi(t),t]$ of a random variable $\xi(t)$. The  average of this functional over all the possible trajectories $\{\xi(t)\}$ in a time-interval $(0,t)$.  is denoted by $\langle R[\xi(t),t]\rangle$. The Shapiro-Loginov formula  reads
\begin{align}
{d\over dt}\langle \xi(t)R[\xi(t),t]\rangle=\langle \xi(t){d\over dt}R[\xi(t),t]\rangle -\lambda\,\langle \xi(t) R[\xi(t),t]\rangle\ ,
\label{shap}
\end{align}
and it is valid for different random processes with exponential correlation function $\langle \xi(t_1) \xi
(t_2)\rangle =\exp [-\lambda |t_1-t_2|]$. We used it in Eq.~(\ref{df}) in the case of telegraph noise,  with $\lambda\equiv\gamma$, for the evaluation of the functional $R[\xi(t),t]\equiv \widetilde{b}_0^{(k)}(t)$, Eq.~(\ref{b0k}).

Averaging Eq. (\ref{b0k}) requires the average of the exponent,
$\exp\big(-i{U\over2}\int_{t'}^t\xi(t'')dt''\big)$. Using the time-ordered Taylor expansion of this exponent, we find
\begin{align}
&\widetilde b_0^{(k)}(t)=\int\limits_0^tK(t,t')dt'\Bigg[1+\sum_{n=1}^\infty \Big(-i{U\over2}\Big)^n\nonumber\\
&\times\int\limits_0^{t'}dt_1\int\limits_0^{t_1}dt_2\cdots
\int\limits_{0}^{t_{n-1}}dt_n\xi(t_1)\xi(t_2)\cdots\xi(t_{n})\Bigg]\ ,
\label{shl}
\end{align}
where $K(t,t')=\exp [(iE_k-\Gamma/2)(t-t')]$.

Let us multiply this equation by $\xi(t)$ and average it over all trajectories, taking into account that $\langle\xi(t)\rangle=0$ (stationary random process). One obtains
\begin{align}
&\langle\xi(t)\widetilde b_0^{(k)}(t)\rangle=\sum_{n=1}^\infty \Big(-i{U\over2}\Big)^n\int\limits_0^tK(t,t')dt'
\nonumber\\
&\times\int\limits_0^{t'}dt_1\int\limits_0^{t_1}dt_2\cdots
\int\limits_{0}^{t_{n-1}}dt_n\langle\xi(t)\xi(t_1)\xi(t_2)
\cdots\xi(t_{n})\rangle\ .
\label{shl2}
\end{align}
The time-derivative of this expression can be written as a sum of three terms,
\begin{align}
{d\over dt}\langle\xi(t)\widetilde b_0^{(k)}(t)\rangle=\sum_{n=1}^\infty\Big(-i{U\over2}\Big)^n
\big(A_1^{(n)}+A_2^{(n)}+A_3^{(n)}\big)\ ,
\label{shl3}
\end{align}
where
\begin{subequations}
\label{shl6}
\begin{align}
&A_1^{(n)}=K(t,t)\int\limits_0^{t}dt_1\cdots
\int\limits_{0}^{t_{n-1}}dt_n\langle\xi(t)\xi(t_1)
\cdots\xi(t_{n})\rangle\ ,\label{shl6a}\\
&A_2^{(n)}=\int\limits_0^t\Big[{d\over dt}K(t,t')\Big]dt'\nonumber\\
&~~~~~~~~~~~~
\times\int\limits_0^{t'}dt_1\cdots
\int\limits_{0}^{t_{n-1}}dt_n\langle\xi(t)\xi(t_1)
\cdots\xi(t_{n})\rangle\ ,\label{shl6b}\\
&A_3^{(n)}=\int\limits_0^tK(t,t')dt'\nonumber\\
&~~~~~~~~~
\times\int\limits_0^{t'}dt_1\cdots
\int\limits_{0}^{t_{n-1}}dt_n{d\over dt}\langle\xi(t)\xi(t_1)
\cdots\xi(t_{n})\rangle\ .
\label{shl6c}
\end{align}
\end{subequations}
Consider now $\langle\xi(t){d\over dt}\widetilde b_0^{(k)}(t)\rangle$. Using Eq.~(\ref{shl}) we easily find
\begin{align}
\langle\xi(t){d\over dt}\widetilde b_0^{(k)}(t)\rangle=\sum_{n=1}^\infty\Big(-i{U\over2}\Big)^n
\big(A_1^{(n)}+A_2^{(n)}\big)\ .
\label{shl4}
\end{align}
Therefore,
\begin{align}
{d\over dt}\langle\xi(t)\widetilde b_0^{(k)}(t)\rangle-\langle\xi(t){d\over dt}\widetilde b_0^{(k)}(t)\rangle=\sum_{n=1}^\infty\Big(-i{U\over2}\Big)^n
A_3^{(n)}\ ,
\end{align}
where $A_3^{(n)}$, given by Eq.~(\ref{shl6c}), contains multiple correlators. For the case $n=2$,   Eq.~(\ref{21}) gives ${d\over dt}\langle\xi(t)\xi(t_1)\rangle =-\gamma\langle\xi(t)\xi(t_1)\rangle$. If the same relation holds for any higher order correlator,
\begin{align}
{d\over dt}\langle\xi(t)\xi(t_1)
\cdots\xi(t_{n})\rangle=-\gamma \langle\xi(t)\xi(t_1)
\cdots\xi(t_{n})\rangle\ ,
\label{shl7}
\end{align}
then it would follow from Eq.~(\ref{shl2}) that
\begin{align}
\sum_{n=1}^\infty\Big(-i{U\over2}\Big)^n
A_3^{(n)}=-\gamma \langle\xi(t)\widetilde b_0^{(k)}(t)\rangle
\label{shl5}
\end{align}
As a result we would arrive to the Shapiro-Loginov formula, Eqs.~(\ref{df}), (\ref{shap}).

Let us now demonstrate that Eq.~(\ref{shl7}) indeed holds for any $n$ in the case of Markovian telegraph process. We do it by extending Eq.~(\ref{excor}) for multiple correlators and using (\ref{17}).
Thus we can write
\begin{align}
&\langle\xi(t)\xi(t_1)\cdots\xi(t_{n})\rangle =\sum_{\xi,\xi_1^{},\ldots,\xi_n^{}}\xi\xi_1^{}\xi_2^{}\cdots\xi_n^{}
p_\xi\nonumber\\
&\times p_{\xi,\xi_1^{}}(t-t_1)p_{\xi_1^{},\xi_2^{}}(t_1-t_2)\cdots p_{\xi_{n-1}^{},\xi_{n}^{}}(t_{n-1}-t_n)
\end{align}
where $\xi,\xi_1^{},\xi_2^{},\ldots,\xi_n^{}=\pm1$,  $p_\xi=1/2$ and
\begin{align}
p_{\xi,\xi_1}^{}(t,t_1)=
\left(\delta_{\xi,\xi_1}-\frac{1}{2}\right)e^{-\gamma |t-t_1|}+\frac{1}{2}\ .
\label{20ap}
\end{align}
Since
\begin{align}
{d\over dt}p_{\xi,\xi_1}^{}(t,t_1)=-\gamma
\left(\delta_{\xi,\xi_1}-\frac{1}{2}\right)e^{-\gamma |t-t_1|}\ ,
\end{align}
we can write
\begin{align}
&{d\over dt}\langle\xi(t)\xi(t_1)
\cdots\xi(t_{n})\rangle=-\gamma \langle\xi(t)\xi(t_1)
\cdots\xi(t_{n})\rangle\nonumber\\
&+{\gamma\over2}\sum_{\xi,\xi_1^{},\ldots,\xi_n^{}}\xi\xi_1^{}
\cdots\xi_n^{}p_\xi\nonumber\\
&~~~~~\times p_{\xi_1^{},\xi_2^{}}(t_1-t_2)\cdots p_{\xi_{n-1}^{},\xi_{n}^{}}(t_{n-1}-t_n)\ .
\label{shl8}
\end{align}
The last term vanishes after summation over $\xi=\pm1$. Then Eq.~(\ref{shl8}) coincides with Eq.~(\ref{shl7}).


\begin{thebibliography}{999}

\bibitem{Imry} S. Datta, {\it Electronic Transport in Mesoscopic Systems}
(Cambridge University Press, New York, 1995);
 Y. Imry, {\it Introduction to Mesoscopic Physics} (Oxford University
Press, Oxford, 1997);
 Y. V. Nazarov and Y. M. Blanter, {\it Quantum Transport:
Introduction to Nanoscience} (Cambridge University Press,
Cambridge, UK, 2009).

\bibitem{fuji} e.g. T. Fujisawa, T. Hayashi, and S. Sasaki, Rep. Prog. Phys.
{\bf 69}, 759 (2006).



\bibitem{1} A.P. Jauho, N.S. Wingreen and Y. Meir, Phys. Rev. B {\bf 50},  5528 (1994); A. Croy and U. Saalmann, Phys. Rev. B {\bf 80}, 245311 (2009).
\bibitem{114} S. A. Gurvitz and Ya. S. Prager, {\it ibid.}
{\bf 53}, 15932 (1996); S. A. Gurvitz, {\it ibid.} {\bf 57}, 6602 (1998).
\bibitem{2} S. Kohler, J. Lehmann and P. H\"anggi, Phys. Rep. {\bf 406}, 379 (2005); V.May and O. K\"uhn, Phys. Rev. B {\bf 77}, 115439 (2008).
\bibitem{weimin} M. W. -Y. Tu and W. -M. Zhang, Phys. Rev. B {\bf 78}, 235311
(2008).
\bibitem{3} S. Kurth, G. Stefanucci, C.O. Almbladh, A. Rubio and E.K.U. Gross, Phys. Rev. B {\bf 72}, 035308 (2005).
\bibitem{4} X. Zheng, F. Wang, C.Y. Yam, Y. Mo and G.H. Chen, Phys. Rev. B {\bf 75}, 195127 (2007); X. Zheng, G.H. Chen, Y. Mo, S. Koo, H. Tian, C. Yam and Y. Yan, J. Chem. Phys. {\bf 133}, 114101 (2010);
Y. Zhang, S. Chen and G.H.Chen, Phys. Rev. B {\bf 87}, 085110 (2013).
\bibitem{5} A. Tikhonov, R.D. Coalson and Y. Dahnovsky, J. Chem. Phys. {\bf 116}, 10909 (2002); L. Arrachea and M. Moskalets, Phys. Rev. B {\bf 74}, 245322 (2006).
\bibitem{avi} A. Schiller and S. Hershfield, Phys. Rev. B {\bf 58} 14978
(1998).
\bibitem{tu} M.W.Y. Tu, A. Aharony, O. Entin-Wohlman, A. Schiller and W.M. Zhang, arXiv:1601.01081.
\bibitem{single} S. Gurvitz, Phys. Scr. {\bf T165}, 014013 (2015).

\bibitem{lan} R. Landauer, Philos. Mag. {\bf 21}, 863 (1970).
\bibitem{sd} S. Dattagupta, {\it Relaxation Phenomena in Condensed Matter Physics} (Academic Press, Orlando, 1987), Chap. VIII.
\bibitem{amnon} A. Aharony, S. Gurvitz, O. Entin-Wohlman, and S. Dattagupta, Phys. Rev. B {\bf 82}, 245417 (2010).


\bibitem{IT} T. Itakura and Y. Tokura, Phys. Rev. B {\bf 67}, 195320 (2003).

\bibitem{13} J. Bergli, Y. M. Galperin, and B. L. Altshuler, New J. Phys. {\bf 11}, 025002 (2009). 
\bibitem{18} B. Cheng, Q. H. Wang, and R. Joynt, Phys. Rev.
A {\bf 78}, 022313 (2008).
\bibitem{alt} Y. M. Galperin, B. L. Altshuler, and D. V. Shantev, in {\it Fundamental
Problems of Mesoscopic Physics}, edited by I. V. Lerner,
B. L. Altshuler, and Y. Gefen, (Kluwer, Dordrecht, 2004), pp.
141-165.

\bibitem{19}  B. Abel and F. Marquardt, Phys.
Rev. B {\bf 78}, 201302 (2008); I. V. Yurkevich, J. Baldwin, I. V.
Lerner, and B. L. Altshuler, {\it ibid.} {\bf 81}, 121305 (2010).

\bibitem{14} S. A. Gurvitz and D. Mozyrsky, Phys. Rev. B {\bf 77}, 075325 (2008).

\bibitem{alt2} E. Paladino, Y. M. Galperin, G. Falci and B. L. Altshuler, Rev. Mod. Phys. {\bf 86}, 361 (2014).

\bibitem{YG} Yu. M. Galperin, N. Zou and K. A. Chao, Phys. Rev. B {\bf 49}, 13728 (1994); Y. M. Galperin and K. A. Chao, Phys. Rev. B {\bf 52}, 12126 (1995).


\bibitem{com1} This equation extends and corrects the initial conditions mentioned after Eq. (9) of Ref. \onlinecite{single}. The correct initial state contains only the states ${\bar l}$ and ${\bar r}$ in the two reservoirs (below the respective Fermi energies), and possibly the state on the dot.



\bibitem{shapiro} V.E. Shapiro and V.M. Loginov, Physica {\bf 91A}, 563 (1978).
\bibitem{klya}  V. Klyatskin, {\em Lectures on Dynamics of Stochastic Systems}, (Elsevier, Amsterdam, 2011).


\bibitem{complex} The imaginary part of such shifts has been used in the literature to generate decoherence. In some cases (e.g. Refs. \onlinecite{schiller,BIH}), the shifts indeed generate decoherence. However, in many other cases the question whether imaginary energy shifts are equivalent to decoherence mechanisms is still a matter of debate, see e.g.
P. W. Brouwer and C. W. J. Beenakker, Phys. Rev. B {\bf 55}, 4695 (1997).

\bibitem{schiller} O. Entin-Wohlman, A. Aharony, Y. Imry, Y. Levinson, and A. Schiller,
Phys. Rev. Lett. {\bf 88}, 166801 (2002).



\bibitem{BIH} A. Aharony, O. Entin-Wohlman, B. I. Halperin and Y. Imry, Phys. Rev. B {\bf 66}, 115311 (2002).

\bibitem{slich} C.P. Slichter, {\em Principles of Magnetic Resonance} (Harper and Row, New York, 1963).
\bibitem{legg} A.J. Leggett, S. Chakravarty, A.T. Dorsey, M.P.A. Fisher, A. Garg, and W. Zwerger, Rev. Mod. Phys. {\bf 59}, 1 (1987).

\bibitem{PWA} J. R. Klauder and P. W. Anderson, Phys. Rev. {\bf 125}, 912 (1962).

\bibitem{bb}
Y.M. Blanter and M. B\"uttiker, Phys. Rep. {\bf 336}, 1 (2000), p.104.
\bibitem{you} J.Q. You, C.H. Lam and H.Z. Zheng, Phys. Rev. B {\bf 62}, 1978 (2000);
G. Stefanucci and C.-O Almbladh, Phys. Rev. B {\bf 69}, 195318 (2004).


\end{thebibliography}
\end{document}